\RequirePackage{fix-cm}

\documentclass[smallextended]{svjour3}       
\usepackage[left=3.17cm,right=3.17cm,top=2.54cm,
headheight=0.5cm,headsep=0.54cm,bottom=2.54cm,footskip=0.79cm
]{geometry}

\usepackage{amsmath,amssymb}

\usepackage{graphicx}
\def\beq{\begin{equation}}
\def\eeq{\end{equation}}
\def\half{\frac{1}{2}}
\def\B{\mathcal{B}}
\def\ket#1{\left|#1\right>}
\begin{document}
	
	\title{Tight upper bound for the maximal expectation value of the Mermin operators}
	
	
	\author{Mohd Asad Siddiqui        \and
		Sk Sazim 
	}
	
	
	\institute{Mohd Asad Siddiqui  \at
		Department of Physics, National Institute of Technology, Patna-800005, India \\
		\email{asad@ctp-jamia.res.in}           
		\emph{Present address: Indian Institute of Science Education and Research Mohali, SAS Nagar, Manauli-140306, India}  
		\and
		Sk Sazim \at
		Quantum Information \& Computation group, Harish-Chandra Research Institute, HBNI, Allahabad, 211019, India\\
		\email{sk.sazimsq49@gmail.com}   
	}

	\date{Received: date / Accepted: date}

	\maketitle

	\begin{abstract}
		The violation of the Mermin inequality (MI) for multipartite quantum states guarantees the existence of nonlocality between either few or all parties. The detection of optimal MI violation is fundamentally important, but current methods only involve numerical optimizations,
		thus hard to find even for three-qubit states. In this paper, we provide a simple and elegant analytical method to achieve the upper bound of Mermin operator for arbitrary three-qubit states. Also, the necessary and sufficient conditions for the tightness of the bound for some class of tri-partite states has been stated.  Finally, we suggest an extension of this result for up to $n$ qubits.

	\end{abstract}

	\section{Introduction}
	
	Quantum nonlocality \cite{nonlocal_rev} has proved to be a  valuable resource in many information processing tasks, e.g. quantum computation \cite{comp_rev}, quantum communication \cite{comm_rev}, quantum key distribution \cite{qkd_rev}, and quantum state estimation \cite{qse_rev}.
	Thus, it's important to mark those states which can reveal nonlocal correlations. The presence of quantum nonlocality between two parties can be revealed using various inequalities -- namely Bell-CHSH inequalities \cite{bell-chsh}.
	
	The optimal quantum violations can be determined using a numerical optimization method, but it gets herculean when we go beyond bipartite cases. Also, the study of nonlocality in the multipartite system is much more profound. For example, in the tripartite system, a new class of nonlocality, a genuine tripartite nonlocality emerges, which can be revealed using Svetlichny inequality (SI) \cite{svet87}. 
	
	Another interesting way to calculate the optimal quantum violation of Bell-CHSH inequality is to use the Horodecki's analytical formula \cite{horo95}. The generalization of this pioneering work for
	the multipartite case is still demanding.  
	 Most of the previous results were obtained doing numerical optimizations, which is a rigorous way to complete the task and even gets harder for mixed states because of stiff optimization conditions. Thus establishing the analytical results for an arbitrary state for the various class of inequalities is much needed.
	    
		The first quest was to build the analytical results for the quantum upper bound in various tripartite inequalities,  viz. the SI and the MI \cite{mer90}. 	 In this direction, Li et al. \cite{li17} have proposed an elegant method for finding the tight upper bound for the Svetlichny operator. 
		 They specifically derived the formula for tripartite states. In this paper, we attempt to extend the Horodecki's result up to $n$ parties, which will be done by employing the Mermin-Klyshko class of inequalities.
		  Note that the MI  \cite{nonlocal_rev} is considered as a natural generalization of Bell-CHSH inequality, and has the advantage of revealing any form of nonlocality present in the system.

	The step towards the search for a quantum upper bound of Mermin operator for three-qubit state has also been attempted earlier in \cite{satya16}, but the obtained bound was far from general and not always tight.  Thus the problem demands further examination for determining the exact optimal bound without going through rigorous optimizations.   
	
	In this work, we manifest the analytical results for arbitrary three-qubit states for obtaining the tight quantum upper bound of Mermin operator. The violation of MI for several three-qubit states such as  GHZ  and  W-states has been computed earlier \cite{merm1,merm11,merm2,merm3} using the numerical optimizations. 
	 Here we propose a method of obtaining the tight expression for the maximum expectation value of Mermin operator, in terms of the singular values of a tripartite correlation matrix.  We also extend our results to  higher parties. This method not only works for pure but also for mixed states. 
	
	The work has been organized as follows. In section-\ref{two}, we obtain the maximum expectation value of the Mermin operator for arbitrary three qubits. We illustrate our results in section-\ref{three}. We then suggest an extension of the results for $n$ qubits in section-\ref{mabknq}. Finally, we conclude in section-\ref{four}.
	
	\section{The Mermin operator and its quantum upper bound}\label{two}
	Let us establish the following correlations using unit vectors $\hat{a}_{i}$, $\hat{b}_{i}$ and $\hat{c}_{i} \in \mathcal{R}^3$ and 
	vector of Pauli spin matrices, $\vec{\sigma}=(\sigma_{1},\sigma_{2},\sigma_{3})$,  
	\begin{eqnarray}
	\mathcal{M}&=&\hat{a}.\vec{\sigma}\otimes
	\hat{b}.\vec{\sigma}\otimes
	\hat{c'}.\vec{\sigma}+\hat{a}.\vec{\sigma}\otimes
	\hat{b'}.\vec{\sigma}\otimes \hat{c}.\vec{\sigma}
	\nonumber\\&+&\hat{a'}.\vec{\sigma}\otimes
	\hat{b}.\vec{\sigma}\otimes
	\hat{c}.\vec{\sigma}-\hat{a'}.\vec{\sigma}\otimes
	\hat{b'}.\vec{\sigma}\otimes
	\hat{c'}.\vec{\sigma}.\label{mermin1}
	\end{eqnarray}
	The operator $\mathcal{M}$ is known as Mermin operator and the MI for three qubits reads $\langle \mathcal{M}\rangle_{\rho}\leq 2$ \cite{mer90,mer90A}. 
	 
	The general three-qubit state $\rho$ in the Hilbert space  $\mathcal{H}=\mathcal{C}^2 \otimes \mathcal{C}^2 \otimes \mathcal{C}^2 $ can be represented as
	\begin{eqnarray}
	\rho=\frac{1}{8}\sum_{\mu,\nu,\gamma=0}^3\Lambda_{\mu\nu\gamma}\sigma_{\mu}\otimes\sigma_{\nu}\otimes \sigma_{\gamma}, \label{3qstate}
	\end{eqnarray}
	where $\Lambda_{\mu\nu\gamma}={\rm Tr}[(\sigma_{\mu}\otimes\sigma_{\nu}\otimes \sigma_{\gamma})\rho]$. The coefficients $\Lambda_{000}=1$ is the normalization condition; 
	$\vec{q}=\{\Lambda_{i00};i=1,2,3\}$, $\vec{r}=\{\Lambda_{0j0};j=1,2,3\}$, $\vec{s}=\{\Lambda_{00k};k=1,2,3\}$ are 
	the Bloch vectors for three parties respectively; $\Theta=[\Lambda_{ij0}]$, $\Phi=[\Lambda_{i0k}]$, $\Omega=[\Lambda_{0jk}]$ are the two party correlation matrices
	and $T=[\Lambda_{ijk}]$ is the tripartite correlation matrix.
	
	The expectation value of the Mermin operator, $\mathcal{M}$ for state $\rho$ is given by
	\begin{eqnarray}
	\langle\mathcal{M}\rangle_{\rho}={\rm Tr}[\mathcal{M} \rho].\label{expectationvalue}
	\end{eqnarray}
	
	\noindent {\bf Theorem}    \textit{The maximum expectation value of Mermin operator for arbitrary three-qubit state $\rho$ is given by 
		$\mathcal{Q}_{\mathcal{M}}:=\max|\langle \mathcal{M}\rangle_{\rho}| \leq  2\sqrt{2}\lambda_{\max}$, where  $\lambda_{\max}$ is the largest singular value of $T$.}
	
	To prove the above theorem we first state the  lemma from \cite{li17}, which will play a central role in proving the above theorem.\\ 
	\noindent {\bf Lemma} \textit{For any vectors $\vec{x}\in\mathcal R^m$ and $\vec{y}\in\mathcal R^n$ and a $m\times n$ rectangular matrix  $\mathcal{B}$,  we have 
		$|\vec{x}^{\intercal} \mathcal{B}\vec{y}|\leq \lambda_{\max}|\vec{x}||\vec{y}|$, where $\lambda_{\max}$ is a largest singular value of $\mathcal{B}$. The equality holds when
		$\vec{x}$ and $\vec{y}$ are the corresponding singular vectors of $\mathcal{B}$ with respect to $\lambda_{\max}$, where $\intercal$ stands for transposition.}\\
	The lemma is a Cauchy-Schwarz inequality. For the detailed proof, we refer readers to \cite{li17}. Let us now prove the above theorem.
	\begin{proof} The expression of Mermin operator in Eq. (\ref{mermin1}) can be recast as  
		\begin{eqnarray}
		\mathcal{M} = \sum_{i,j,k}\left(a_{i} b_{j} c'_{k} + a_{i} b'_{j} c_{k}+ a'_{i} b_{j} c_{k}-a'_{i} b'_{j} c'_{k}\right)
		\sigma_{i}\otimes\sigma_{j}\otimes\sigma_{k} \nonumber    
		\end{eqnarray}
		The expectation of Mermin operator is thus given by
		\begin{eqnarray}
		\langle \mathcal{M}\rangle_\rho &=& \sum_{i,j,k}\left(a_{i} b_{j} c'_{k} + a_{i} b'_{j} c_{k}+ a'_{i} b_{j} c_{k}-a'_{i} b'_{j} c'_{k}\right) {\rm Tr}\left[\rho(\sigma_{i}\otimes\sigma_{j}\otimes\sigma_{k})\right] \nonumber     \\
		&=& \sum_{i,j,k}\left[a_{i} b_{j} c'_{k} + a_{i} b'_{j} c_{k}+ a'_{i} b_{j} c_{k}-a'_{i} b'_{j} c'_{k}\right] \Lambda_{ijk}\nonumber \\
		&=& \hat{a}^{\intercal} T (\hat{b}\otimes \hat{c}'+\hat{b}'\otimes \hat{c})+\hat{a}'^{\intercal} T(\hat{b}\otimes \hat{c}-\hat{b}'\otimes \hat{c}').
		\label{mermin2}
		\end{eqnarray}
		Let $\theta_b$ and $\theta_c$ are the angles  between $\hat{b}$ and $\hat{b}'$, and $\hat{c}$ and $\hat{c}'$ respectively, then 
		\begin{eqnarray*}
			|\hat{b}\otimes \hat{c}'+\hat{b}'\otimes \hat{c}|^2=2+2\langle \hat{b}, \hat{b}'\rangle\langle \hat{c}, \hat{c}'\rangle=2+2\cos\theta_b\cos\theta_c,\\
			|\hat{b}\otimes \hat{c}-\hat{b}'\otimes \hat{c}'|^2=2-2\langle \hat{b}, \hat{b}'\rangle\langle \hat{c}, \hat{c}'\rangle=2-2\cos\theta_b\cos\theta_c.
		\end{eqnarray*}
		Consider $\theta_{bc}$ to be the principal angle such that $\cos\theta_b\cos\theta_c=\cos\theta_{bc}$. Hence,
		\begin{eqnarray}
		|\hat{b}\otimes \hat{c}'+\hat{b}'\otimes \hat{c}|^2&=&4\cos^2\frac{\theta_{bc}}2,\nonumber\\
		|\hat{b}\otimes \hat{c}-\hat{b}'\otimes \hat{c}'|^2&=&4\sin^2\frac{\theta_{bc}}2.
		\label{trig_id}
		\end{eqnarray}
		Using the Lemma in Eq. (\ref{mermin2}) and employing Eq. (\ref{trig_id}), we get 
		\begin{eqnarray}
		|\langle \mathcal{M}\rangle_\rho| &\leq &
		\lambda_{\max}\left(|\hat{a}||\hat{b}\otimes \hat{c}'+\hat{b}'\otimes \hat{c}|+|\hat{a}'| |\hat{b}\otimes \hat{c}-\hat{b}'\otimes \hat{c}'| \right)\nonumber\\
		&=&2\lambda_{\max}\left(\cos\frac{\theta_{bc}}2+\sin\frac{\theta_{bc}}2\right)\nonumber\\
		&\leq & 2\sqrt{2}\lambda_{\max}
		\label{mi_ineq}
		\end{eqnarray}
		The above inequality is achieved using $x\cos\theta+y\sin\theta\leq \sqrt{x^2+y^2}$, equality holds for $\tan\theta=y/x.$ \end{proof}
	
	{\em Tightness of the bound}.- Let us now discuss the conditions where the upper bound (\ref{mi_ineq}) saturates. To saturate the upper bound in the Theorem, one can choose $\theta_{bc}=\frac{\pi}{2}$. 
	Also, the Lemma tells us that the first inequality in Eq. (\ref{mi_ineq}) saturates if the degeneracy of $\lambda_{\max}$ is more than one, and corresponding to 
	$\lambda_{\max}$, there exist two nine-dimensional singular vectors of the form $\hat{b}\otimes \hat{c}'+\hat{b}'\otimes \hat{c}$ and 
	$\hat{b}\otimes \hat{c}-\hat{b}'\otimes \hat{c}'$, respectively.

	\section{Illustrations of the upper bound of the Mermin operator}\label{three}
	Here we will discuss the quantum upper bound of Mermin operator for some class of states and comment on whether the bound saturates. 
	
	{\em Example 1}.- The GHZ-symmetric states of three qubits can be expressed as 
	\begin{eqnarray}\label{ghz1}
	\rho(\ell, \vartheta) &=&\frac{1}{8}(1- \frac{4\vartheta}{\sqrt{3}})\mathbb{I}_8
	+(\frac{2\vartheta}{\sqrt{3}}+ \ell)|GHZ_{+}\rangle\langle GHZ_{+}| +
	(\frac{2\vartheta}{\sqrt{3}}- \ell)|GHZ_{-}\rangle\langle GHZ_{-}|,
	\end{eqnarray}
	where $|GHZ_{\pm}\rangle = \frac{1}{\sqrt{2}}(|000\rangle \pm |111\rangle)$ and $\mathbb{I}_8$ is the identity matrix of order $8$. The matrix, 
	$\rho(\ell, \vartheta)$ will be positive semidefinite iff 
	\begin{equation}\label{c1}
	-\frac{1}{4\sqrt{3}}\leq \vartheta \leq \frac{\sqrt{3}}{4}\hspace{0.2cm}\mbox{and}\hspace{0.2cm}|\ell| \leq \frac{1}{8}+\frac{\sqrt{3}}{2} \vartheta.
	\end{equation}
	This family of density matrices creates a triangle in the state space and contains the GHZ states, some W-type states and the maximally mixed state $\frac{1}{8}\mathbb{I}$. 
	The generalized three-qubit Werner states 
	are oriented along the line, $\vartheta=\frac{\sqrt{3}}{2}\ell$. We refer readers to reference \cite{ghz_symm1} for the detailed geometric picture of the GHZ-symmetric states for three qubits.
	
	The tripartite correlation matrix, $T$ for the state $\rho(\ell, \vartheta)$ is given by
	\begin{eqnarray*}
		&&T=2\ell\begin{pmatrix}
			1 & 0 & 0 & 0 & -1 & 0 & 0 & 0 & 0 \\
			0 & -1 & 0 & -1 & 0 & 0 & 0 & 0 & 0 \\
			0 & 0 & 0 & 0 & 0 & 0 & 0 & 0 & 0
		\end{pmatrix}
		\label{ghz-sym}
	\end{eqnarray*}
	
	which has one non-zero singular value, $\lambda_{\max}= 2\sqrt{2} |\ell|$ with degeneracy $2$. The corresponding singular vectors are $\{0,-1,0,-1,0,0,0,0,0\}$ and 
	$\{1,0,0,0,-1,0,0,0,0\}$ which can be decomposed into $(\{1,0,0\}\otimes\{0,-1,0\}+\{0,-1,0\}\otimes\{1,0,0\})^{\intercal}$ and 
	$(\{1,0,0\}\otimes\{1,0,0\}-\{0,-1,0\}\otimes\{0,-1,0\})^{\intercal}$ respectively. Therefore, the angle $\theta_{ac}$ comes out to be $\frac{\pi}{2}$. Hence, with these 
	settings the inequality Eq. (\ref{mi_ineq}) becomes equality, and thus the upper bound saturates for $\rho(\ell, \vartheta)$.
	
	The maximum expectation value of Mermin operator for the GHZ-symmetric state is 
	\begin{eqnarray}
	\mathcal{Q}_{\mathcal{M}}=8 |\ell|. \label{max1}
	\end{eqnarray}
	This value of Mermin operator has been found in \cite{MI-SI_ent} using numerical methods. Hence, the GHZ-symmetric state will 
	violate MI if $|\ell|>\frac{1}{4}$.
	
	One way of computing multipartite entanglement is to compute genuine multipartite concurrence \cite{gmc1,gmc2}. There exist a closed form of 
	genuine multipartite concurrence for X-states \cite{gmcx} (see Appendix \ref{appen}). 
	We can easily show that the maximum quantum violation of Mermin operator for GHZ symmetric state is related with its entanglement as captured by 
	genuine multipartite concurrence ($C_m$) \cite{MI-SI_ent},
	\begin{equation*}
	\mathcal{Q}_{\mathcal{M}}=4(C_m+\frac{3}{4}-\sqrt{3}\vartheta).
	\end{equation*}
	Hence, the GHZ symmetric state will violate MI if $C_m>\sqrt{3}\vartheta-\frac{1}{4}$. 
	
	{\em Example 2}.- Consider the state introduced in \cite{new_ghz} for three qubit scenario, 
	\begin{equation}
	\tilde{\rho}_{GHZ}=p|GHZ\rangle\langle GHZ|+\frac{1-p}{4}\mathbb{I}_2\otimes \tilde{\mathbb{I}},
	\label{mix34}
	\end{equation}
	where $\tilde{\mathbb{I}}={\rm diag}(1,0,0,1)$. Note that $|GHZ\rangle\equiv|GHZ_{+}\rangle$. 
	
	The tripartite correlation matrix, $T$ for the state $\tilde{\rho}_{GHZ}$ is  
	\begin{eqnarray*}
		&&T=p\begin{pmatrix}
			1 & 0 & 0 & 0 & -1 & 0 & 0 & 0 & 0\\
			0 & -1 & 0 & -1 & 0 & 0 & 0 & 0 & 0\\
			0 & 0 & 0 & 0 & 0 & 0 & 0 & 0 & 0
		\end{pmatrix}
		\label{ghzwh}
	\end{eqnarray*}
	The only non-zero singular value of $T$ is $\sqrt{2}p$ and it has degeneracy two. 
	Using the above argument of { \em Example-1}, one can show that the quantum upper bound of Mermin operator for the state $\tilde{\rho}_{GHZ}$ is indeed tight. 
	The maximum expectation value of Mermin operator for the above state is
	\begin{eqnarray*}
		\mathcal{Q}_{\mathcal{M}}=4 p. \label{maxia}
	\end{eqnarray*}
	Hence, the state $\tilde{\rho}_{GHZ}$ violates MI if $p>\frac{1}{2}$. For $p=1$, the value for Mermin operator is $4$ which is the maximum possible value 
	as predicted in the literature \cite{nonlocal_rev}.
	
	The connection between the entanglement and the optimal value of Mermin operator for this class of states can be expressed as 
	$\mathcal{Q}_{\mathcal{M}}=\frac{4}{3}(2C_m+1)$. Hence, this state will violate MI if $C_m>\frac{1}{4}$.
	
\begin{figure}[t]
		\centering
		\includegraphics[scale=0.9]{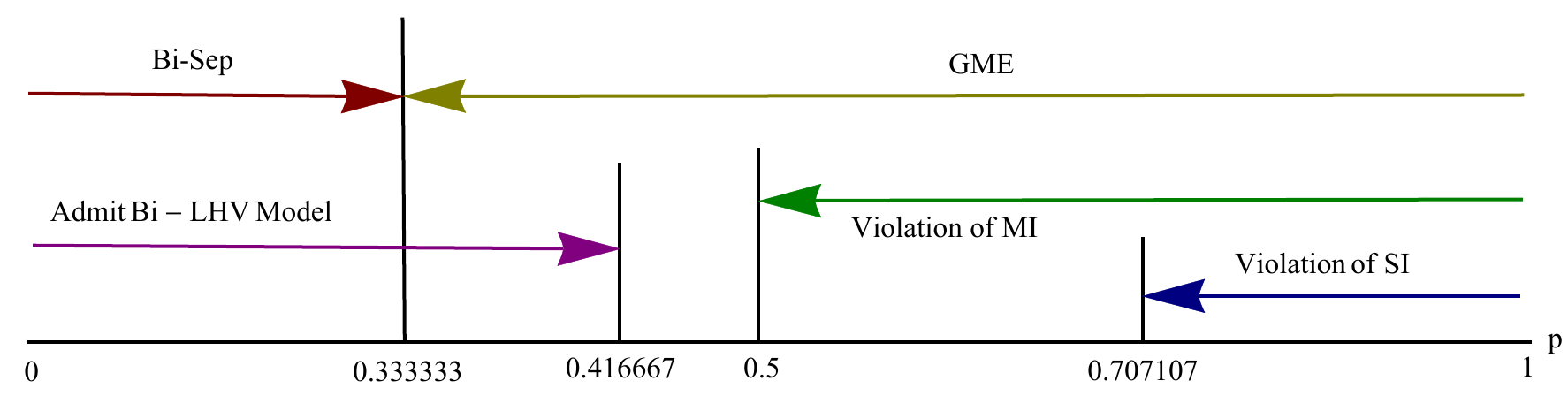}
		\caption{(Color online) {\em Schematic diagram of nonlocality and entanglement for the state, $\tilde{\rho}_{GHZ}$}. Due to \cite{new_ghz}, 
			the state $\tilde{\rho}_{GHZ}$ admits bi-local hidden variable model for $0\leq p \leq \frac{5}{12}$ and is genuinely multipartite entangled for 
			$\frac{1}{3}\leq p\leq 1$. In Ref. \cite{li17}, it has  been shown that the state will violate the SI if $\frac{1}{\sqrt{2}}< p\leq 1$. Whereas our results 
			point out that the state will violate the MI if $\frac{1}{2}< p\leq 1$.}\label{colornoise}

	\end{figure}

	In Fig. \ref{colornoise}, we highlight the violation of MI  along with earlier existing bounds on nonlocality and entanglement for the state $\tilde{\rho}_{GHZ}$. Our findings show that we might be able to detect more entanglement using the violation of MI.
	
	The similar analysis goes for the state, $\rho_{GHZ}=p|GHZ\rangle\langle GHZ|+\frac{1-p}{8}\mathbb{I}_8$. The optimal quantum value of Mermin operator for this state is  also
	$\mathcal{Q}_{\mathcal{M}}=4 p$. The relation between the genuine multipartite concurrence and optimal quantum value for this state is 
	$\mathcal{Q}_{\mathcal{M}}=\frac{4}{7}(4C_m+3)$. Hence,  the state $\rho_{GHZ}$ will violate MI if $C_m>\frac{1}{8}$.
	
	Although all the above-discussed class of states has tight quantum bound for Mermin operator, the following state does not admit such vintage.
	
	\vspace{2cm}
	{\em Example 3}.- A mixed W state can be expressed as,
	\begin{equation*}
	\rho_{W}=p|W\rangle\langle W|+\frac{1-p}{8}\mathbb{I}_8,
	\end{equation*}
	where $|W\rangle=\frac{1}{\sqrt{3}}(|001\rangle + |010\rangle + |100\rangle)$. The correlation matrix, $T$ for the state $\rho_W$, can therefore 
	be written as 
	\begin{eqnarray*}
		&&T=p\begin{pmatrix}
			0 & 0 & \frac{2}{3} & 0 & 0 & 0 & \frac{2}{3} & 0 & 0\\
			0 & 0 & 0 & 0 & 0 & \frac{2}{3} & 0 & \frac{2}{3}& 0\\
			\frac{2}{3} & 0 & 0 & 0 & \frac{2}{3} & 0 & 0 & 0 & -1
		\end{pmatrix}
	\end{eqnarray*}
	The singular values of $T$ are $\frac{\sqrt{17}}{3}p$ with degeneracy one and $\frac{2\sqrt{2}}{3}p$ with degeneracy two. The quantum upper bound of Mermin operator for 
	the state $\rho_{W}$ comes out to be 
	\begin{eqnarray*}
		\mathcal{Q}_{\mathcal{M}}\leq \frac{2\sqrt{34}}{3}p
	\end{eqnarray*} 
	For $p=1$, the upper bound is $\mathcal{Q}_{\mathcal{M}}\lesssim 3.8873$ which might not be tight \cite{cab02}.
	
	\section{Extension to N qubit states}\label{mabknq}
	
	The Mermin-Ardehali-Belinskii-Klyshko (MABK) operator \cite{mer90,ard92,mer90B} for $n$-qubit state is given by:
	\beq
	\B_n=\B_{n-1}\otimes\half(\hat a_n.\vec\sigma+\hat a_n'.\vec\sigma)
	+\B_{n-1}'\otimes\half(\hat a_n.\vec\sigma-\hat a_n'.\vec\sigma),
	\label{mabk}
	\eeq
	The optimal bound of (\ref{mabk}) will be $\B_n^2\le2^{n+1}.$\\
	
	An arbitrary $n$-qubit state $\varrho$ in Hilbert space  $\mathcal{H}=\mathcal{C}_1^2 \otimes ... 
	\otimes \mathcal{C}_n^2 $ can be written as
	\begin{equation}
	\varrho=\frac{1}{2^n} \sum_{\mu_1,...,\mu_n=0}^{3} \Lambda_{\mu_1...\mu_n} \mbox{ }
	\sigma_{\mu_1} \otimes ... \otimes \sigma_{\mu_n}, \label{nstate}
	\end{equation}
	where $\sigma_0$ is identity operator in the $n$-qubit Hilbert space, and $\sigma_{\mu_n}$ are Pauli operators for
	three orthogonal directions $\mu_n=1,2,3$. The set of real coefficients  
	$\Lambda_{\mu_1...\mu_n}(={\rm Tr}[\sigma_{\mu_1} \otimes ... \otimes \sigma_{\mu_n}\varrho])$ 
	gives the correlation matrix $T=[\Lambda_{\mu_1...\mu_n}]$.
	\\
	
	\noindent {\bf Hypothesis:} \textit{The optimal quantum value of MABK operator for arbitrary n-qubit state $\varrho$ is given by 
		$\mathcal{Q}_{\mathcal{B}_n}:=\max|\langle \mathcal{B}_n\rangle_{\varrho}| \leq  2\sqrt{2}\lambda_{\max}$, where  $\lambda_{\max}$ is the largest singular value of $T$.}
	\\
	
The above hypothesis is based on the results of the four-qubit MABK operator,  given by

	%
	\begin{eqnarray}
	\mathcal{B}_4 &=&(\hat{a}.\vec{\sigma}\otimes
	\hat{b}.\vec{\sigma} -\hat{a'}.\vec{\sigma}\otimes
	\hat{b'}.\vec{\sigma})\otimes
	[\hat{c'}.\vec{\sigma}\otimes \half{(\hat{d}+\hat{d'}).\vec{\sigma}}
	-\hat{c}.\vec{\sigma}\otimes \half (\hat{d}-\hat{d'}).\vec{\sigma}] \nonumber \\ 
	&+& (\hat{a}.\vec{\sigma}\otimes
	\hat{b'}.\vec{\sigma} +\hat{a'}.\vec{\sigma}\otimes
	\hat{b}.\vec{\sigma})\otimes
	[\hat{c}.\vec{\sigma}\otimes \half{(\hat{d}+\hat{d'}).\vec{\sigma}}
	+\hat{c'}.\vec{\sigma}\otimes \half (\hat{d}-\hat{d'}).\vec{\sigma}],\label{4qubit}
	\end{eqnarray}
	
	The expectation value is given by
	\begin{eqnarray}
	\langle \mathcal{B}_4\rangle_\varrho &=& \sum_{i,j,k,l}\Big[(a_{i} b_{j}-a'_{i} b'_{j}) [c'_{k} \frac{({d_{l}}+{d'_{l}})} {2}
	-c_{k} \frac{({d_{l}}-{d'_{l}})} {2}]+
	(a_{i} b'_{j}+a'_{i} b_{j})  [c_{k} \frac{({d_{l}}+{d'_{l}})} {2}
	+c'_{k} \frac{({d_{l}}-{d'_{l}})} {2}]\Big] \nonumber\\ &\times&{\rm Tr}\left[\rho(\sigma_{i}\otimes\sigma_{j}\otimes\sigma_{k}
	\otimes\sigma_{l})\right] \nonumber     \\
	&=& \sum_{i,j,k,l}\Big[(a_{i} b_{j}-a'_{i} b'_{j}) [c'_{k} \frac{({d_{l}}+{d'_{l}})} {2}
	-c_{k} \frac{({d_{l}}-{d'_{l}})} {2}]+
	(a_{i} b'_{j}+a'_{i} b_{j}) [c_{k} \frac{({d_{l}}+{d'_{l}})} {2}
	+c'_{k} \frac{({d_{l}}-{d'_{l}})} {2}]\Big] \Lambda_{ijkl}\nonumber \\
	&=& (\hat{a}\otimes \hat{b}-\hat{a}'\otimes \hat{b}')^{\intercal} T  (\hat{c}'\otimes \frac{(\hat{d}+\hat{d'})} {2}
	-\hat{c}\otimes \frac{(\hat{d}-\hat{d'})} {2})+(\hat{a}\otimes \hat{b}'+\hat{a}'\otimes \hat{b})^{\intercal} T  (\hat{c}\otimes \frac{(\hat{d}+\hat{d'})} {2}
	+\hat{c}'\otimes \frac{(\hat{d}-\hat{d'})} {2}),\nonumber\\
	\label{mermin4}
	\end{eqnarray}
	
	Using the Lemma in above equation, we get
	\begin{eqnarray}
	|\langle \mathcal{B}_4\rangle_\varrho| &\leq &
	\lambda_{\max}\Big(|(\hat{a}\otimes \hat{b}-\hat{a}'\otimes \hat{b}')||(\hat{c}'\otimes \frac{(\hat{d}+\hat{d'})} {2}
	-\hat{c}\otimes \frac{(\hat{d}-\hat{d'})} {2})|+|(\hat{a}\otimes \hat{b}'+\hat{a}'\otimes \hat{b})| \nonumber\\&\times&| (\hat{c}\otimes \frac{(\hat{d}+\hat{d'})} {2}
	+\hat{c}'\otimes \frac{(\hat{d}-\hat{d'})} {2}| \Big).
	\label{mi_e}
	\end{eqnarray}
	
	Taking $\theta_a$,  $\theta_b$,  $\theta_c$, and $\theta_d$ as the angles between $\hat{a}$ and $\hat{a}'$; $\hat{b}$ and $\hat{b}'$; $\hat{c}$ and $\hat{c}'$; $\hat{d}$ and $\hat{d}'$ respectively, we get 
	\begin{eqnarray}
	|\hat{a}\otimes \hat{b}-\hat{a}'\otimes \hat{b}'|^2=2-2\langle \hat{a}, \hat{a}'\rangle\langle \hat{b}, \hat{b}'\rangle&=&2-2\cos\theta_a\cos\theta_b, \nonumber\\
	|\hat{a}\otimes \hat{b}'+\hat{a}'\otimes \hat{b}|^2=2+2\langle \hat{a}, \hat{a}'\rangle\langle \hat{b}, \hat{b}'\rangle&=&2+2\cos\theta_a\cos\theta_b, \nonumber\\
	|\hat{c}'\otimes \frac{(\hat{d}+\hat{d'})} {2}
	-\hat{c}\otimes \frac{(\hat{d}-\hat{d'})} {2}|^2&=&1, \nonumber\\
	|\hat{c}\otimes \frac{(\hat{d}+\hat{d'})} {2}
	+\hat{c}'\otimes \frac{(\hat{d}-\hat{d'})} {2}|^2&=&1. 
	\label{rig}
	\end{eqnarray}
	Consider $\theta_{ab}$ to be the principal angle such that $\cos\theta_a\cos\theta_b=\cos\theta_{ab}$. Hence,
	\begin{eqnarray}
	|\hat{a}\otimes \hat{b}-\hat{a}'\otimes \hat{b}'|^2&=&4\sin^2\frac{\theta_{ab}}2,  \nonumber\\
	|\hat{a}\otimes \hat{b}'+\hat{a}'\otimes \hat{b}|^2&=&4\cos^2\frac{\theta_{ab}}2,\nonumber\\
	\label{trig_i}
	\end{eqnarray}
	Using Eq. (\ref{rig}) and Eq. (\ref{trig_i})  in Eq. (\ref{mi_e}), we get 
	\begin{eqnarray}
	|\langle \mathcal{B}_4\rangle_\varrho| &\leq &
	2\lambda_{\max}\left(\sin\frac{\theta_{ab}}2+\cos\frac{\theta_{ab}}2\right)\nonumber\\
	&\leq & 2\sqrt{2}\lambda_{\max}
	\label{mi_ine}
	\end{eqnarray}
	The above inequality is achieved using $x\cos\theta+y\sin\theta\leq \sqrt{x^2+y^2}$, equality holds for $\tan\theta=y/x.$ \par

	The tightness condition of Eq. (\ref{mi_ine})  will be similar to a three-qubit case. Here, the first inequality becomes equality when we choose $\theta_{ab}=\frac{\pi}{2}$ . 
	 Also, the Lemma tells us that the Eq. (\ref{mi_e}) saturates if the degeneracy of $\lambda_{\max}$ is more than one, and corresponding to $\lambda_{\max}$, there exist two nine-dimensional singular vectors of the form $\hat{a}\otimes \hat{b} - \hat{a}'\otimes \hat{b}'$ and 
	 $\hat{a}\otimes \hat{b}' + \hat{a}'\otimes \hat{b}$, respectively.

	As an example, we consider a class of four-qubit  GHZ  state $(\cos{\phi}\ket{0000}+\sin{\phi}\ket{1111})$. Its optimal quantum violation will be given by $\mathcal{Q}_{\mathcal{B}_4} = 2 \sqrt{2} \max{(1, 2 \sin{2 \phi})}$. 
	The optimal bound for GHZ state ($\phi=\pi/4$) comes out to be $4\sqrt{2}$, maximal for any four-qubit 
	MABK operator.
	
	 We feel that the proposed method can also work for obtaining the quantum upper bound for MABK operator beyond four parties. Thus we suggest an extension of our results which might be helpful in obtaining the optimal quantum bound of MABK operator for an arbitrary $n$-qubit state.
	

	\section{Conclusions and Discussion}\label{four}
	Foundational interest in nonlocality is reinforced with the development of quantum information and computation theory.

	In this work, we proffer a quantitative analysis of MI for three qubits by computing the maximum expectation value of the Mermin operator analytically. The obtained maximal value is shown to be tight for the state which produces the correlation matrix with more than one degenerate largest singular values. Also, the maximal quantum bound for several noisy states are shown to be tight. The proposed method not only works for pure states but also for mixed. As MI is very useful in detecting any form of nonlocality in multipartite systems, our method will open a new pathway. We have also discussed the possible relation between multipartite entanglement and the maximum expectation value of Mermin operator for mixed states which might help in experimental characterization of multipartite entanglement. An extension to n-qubit systems has also been suggested for MABK inequality, which tells us that this approach can be useful for multiparty inequalities. This might work as well for other facet inequalities \cite{facet,facet1}.
	
	We hope that our findings will help in understanding the complex structure of nonlocality in multipartite systems.
	
	{\em Acknowledgements}.-  MAS thanks Dr. Sibasish Ghosh for fruitful discussions on multipartite nonlocality during his IMSc visit. 
	He acknowledges the support received from Ramanujan Fellowship research grant (SB/S2/RJN-083/2014) and from HRI during his visits.

	\appendix
	\section{Genuine multipartite concurrence }\label{appen}
	Let us briefly review the genuine multipartite concurrence $(C_{m})$, used in our paper.
	
	The  $C_{m}$ is an important tool to distinguish the biseparable states from  multi-partite entangled states. It is a monotone of 
	genuinely multipartite entanglement which is convex, invariant under local unitary transformations, and non-increasing under local 
	operations and classical communications (LOCC) \cite{gmc1}. Its measure is given by $C_{m}(|\psi\rangle):= \textmd{min}_j \sqrt{2 (1-\mathcal{P}_j(|\psi\rangle))}$, 
	where  $\mathcal{P}_j(|\psi\rangle)$ is the purity of the $j^{th}$ bipartition of $|\psi\rangle$. There exists a closed form of 
	genuine multipartite concurrence for the three-qubit X-states, $\rho_X$ of the form 
	\begin{eqnarray*}
		&&\rho_X=\begin{pmatrix}
			M &  Z  \\
			Z^{\dagger}   & N 
		\end{pmatrix},
	\end{eqnarray*}
	where $M={\rm diag}(m_1,m_2,m_3,m_4)$, $N={\rm diag}(n_4,n_3,n_2,n_1)$ and $Z={\rm antidiag}(z_4,z_3,z_2,z_1)$. The $\dagger$ denotes the complex conjugation. 
	The $C_{m}$ for the above state is given as \cite{gmcx}
	\begin{equation}\label{4v}
	C_{m}=2\,\textmd{max}_i\{0,|z_i|-w_i\}
	\end{equation}
	where $w_i=\sum_{j\neq i}\sqrt{m_jn_j}$.

\end{document}